\documentclass[doublecol,linenumbers]{epl2}

\usepackage{graphicx}
\usepackage{epsfig}
\usepackage{dcolumn}
\usepackage{bm}
\usepackage{color}
\usepackage{SIunits}
\usepackage{epstopdf}
 \usepackage{units}
\usepackage{ulem}
\usepackage{times,mathptm}
\usepackage{textcomp}
\usepackage{soul}
\newcommand\al{\textit{et~al.}\ }

\def\mum{\nobreak\mbox{$\;$\textnormal{\textmu m}}}

\def\strutdepth{\dp\strutbox}
\def\nw#1{\strut\vadjust{\kern-\strutdepth\vtop to0pt{\vss\hbox to\hsize
{\hskip\hsize\hskip5pt$\leftarrow$\hss\strut}}}{\em #1}}

\title{Single particles accelerate final stages of capillary break up}

\author{Anke~Lindner\inst{1}, Jorge Eduardo~Fiscina\inst{2,3}, Christian~Wagner\inst{3}}
\shortauthor{A. Lindner \etal}

\institute{
\inst{1} PMMH-ESPCI, 10, rue Vauquelin, 75231 Paris Cedex 05, France.\\
\inst{2} Gravitation Group, TATA institute of fundamental research, 1 Homi Bhabha Rd., 400005 Mumbai, India.\\
\inst{3}Experimentalphysik, Universit\"at des Saarlandes, Postfach 151150,
66041 Saarbr\"ucken, Germany.}

\abstract{
Droplet formation of suspensions is present in many industrial and technological processes such as coating and food engineering. Whilst the finite time singularity of the minimum neck diameter in capillary break-up of simple liquids can be described by well known self-similarity solutions, the pinching of non-Brownian suspension depends in a complex way on the particle dynamics in the thinning thread. Here we focus on the very dilute regime where the filament contains only isolated beads to identify the physical mechanisms leading to the pronounced  acceleration of the filament thinning observed. This accelerated regime is characterized by an asymmetric shape of the filament with an enhanced curvature that depends on the size and the spatial distribution of the particles within the capillary thread.
}


\begin{document}

\maketitle

\begin{figure}[h!]
	\centering
		\includegraphics[width=0.85\linewidth]{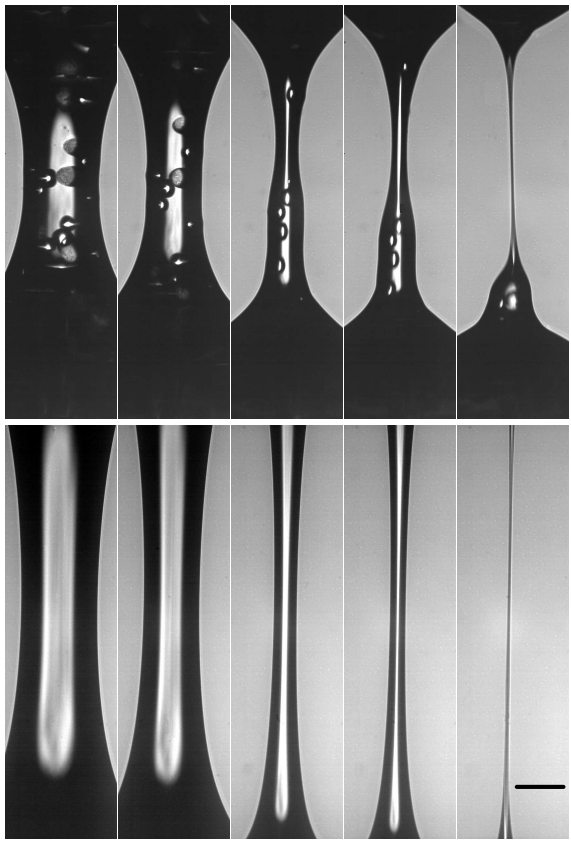}
	\caption{High speed images of capillary break-up of a $3$ wt\% suspension of $80\mum$ particles (top sequence) and silicon oil (bottom sequence). Frames are taken at the same minimum neck diameters, i.e. at times before pinch-off $t=\tau_p-\tau$ = $17.53$, $12.33$, $5.19$, $3.90$, and $1.29$ ms for the suspension and  $t$ =  $19.48$ , $15.58$, $7.79$, $5.84$, and $1.94$ ms for the oil before pinch-off. The scale bar is $0.2$ mm.}
	\label{frames}
\end{figure}

Droplets are present in our daily life and their formation has been investigated scientifically since many decades or even centuries. Yet, the physical mechanisms of capillary break-up of even simple fluids has been only understood within the last 20 years \cite{Eggers1993,Papageorgiou1995}. The final stages of break up are governed by self similar laws that depend only on the fluid parameters such as surface tension, viscosity and density \cite{EV08}. The situation is less clear for complex fluids such as polymer solutions \cite{Amarouchene2001,Wagner2005,Sattler2008,Sattler2012} or suspensions \cite{Bischoff,Chellamuthu,Smith, Roche2011}. Suspensions can be found in many industrial applications, such as painting, food processing or cosmetics and a better understanding of the capillary break up process will be important for a better control of e.g. the dosage.

Non-Brownian suspensions have recently attracted interest as their detachment dynamics are function of the individual and collective particle dynamics in the thread and cannot be described using an effective fluid approach only. Several distinct regimes have been identified during suspension detachment \cite{Furbank2004, Furbank2007, Bonnoit12, Miskin2012} and in particular, an acceleration of the detachment compared to the interstitial fluid has been observed during the very late stages of droplet detachment \cite{vanDeen13}.  Numerical simulations \cite{McIlroy2014} have confirmed these experimental findings and have also indicated the exitance of an accelerated regime where the thinning rate is faster than the thinning rate of the pure solvent. While this acceleration has been experimentally quantified as a function of the number of particles in the thread \cite{vanDeen13} the exact thinning dynamics in the accelerated regime were so far outside the experimental resolution. Here we present measurements on dilute suspensions where only very few particles or even no particles at all are present in the thinning capillary bridge. At these small concentrations the viscosity of the suspensions can be considered to remain equal to the viscosity of the suspending fluid and the change in thinning dynamics can unambiguously be attributed to the presence of single beads in the thread. High speed imaging in combination with a "super resolution technique" \cite{Super} are used to characterize the break up dynamics (Fig. \ref{frames}).

A capillary bridge of sufficient length is prone to the Rayleigh Plateau instability. This is also true for a droplet hanging from a faucet \cite{Sattler2008,Rehberg},  and the capillary neck first thins exponentially in time with a characteristic time constant that depends on the diameter of the neck and the fluid parameters.  Close to the break up, the dynamics are described by self similar laws, and depending on the fluid parameters, one first observes a regime that is governed by viscosity (the Stokes regime) and then by viscosity and inertia (the Navier-Stokes regime) \cite{Rehberg}. Recently, it was shown that in the transition between these two regimes a short lasting pure inertial regime might be identified, and of course, for other fluid parameters, different sequences can be observed \cite{Barsan2015}. Here we show that the presence of even single particles in or near the capillary neck strongly perturbs the thinning dynamics and leads to a new regime with an accelerated thinning of the filament. Very close to pinch-off, the thinning dynamics eventually return to the self-similar Navier-Stokes regime, when the filament is fully depleted of beads. We quantify for the first time the transition points and the velocity of this accelerated regime and quantitatively show that they depend on the size and the spatial distributions of particles in the capillary neck.

\begin{figure}[h!]
	\centering
		\includegraphics[width=1.0\linewidth]{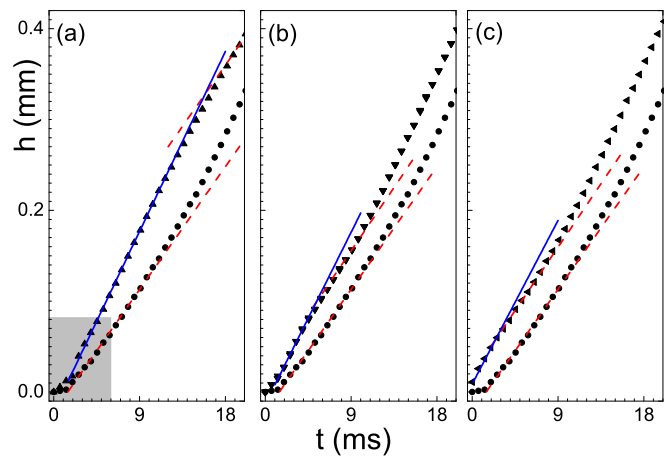}
	\caption{(Color online) The minimum neck diameter $h(t)$ of the thinning capillary bridge of the silicon oil (circles) and the suspensions (triangles). The data shown for the silicon oil is identical on all three panels. From left to right: $80 \mum (\phi=3\%)$, $40 \mum (\phi=1\%)$ and $20 \mum (\phi=3\%)$. The red lines have a fixed slope of $0.015 ms^{-1}$ and indicate the Stokes regime. The blue lines have been fitted with a linear law and indicate the accelerated regime (see also Fig. \ref{slope}). The thinning dynamics close to pinch-off for the area indicated by the square on panel (a) are shown in Fig. \ref{self similar} using experiments with a higher resolution.}	
\label{case studies}
\end{figure}

\begin{figure}[htbp]
	\centering
		\includegraphics[width=0.7\linewidth]{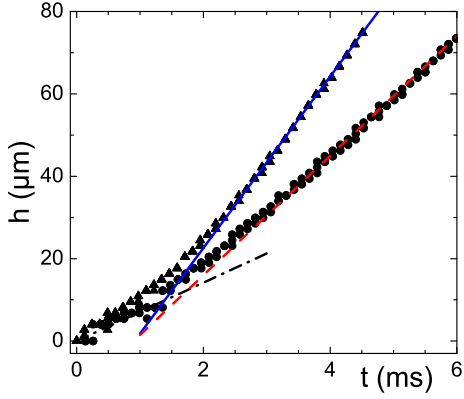}
	\caption{(Color online) The minimum neck diameter $h(t)$ of the thinning capillary bridge of the silicon oil (circles) and the $80 \mum$ suspension (triangles) corresponding to the results within the black square in Fig. \ref{case studies} (a) obtained from experiments with a higher resolution. Lines with fixed slopes of $0.007 ms^{-1}$ (black) and $0.015 ms^{-1}$ (red) illustrate the transition from the Stokes to the Navier-Stokes regime for the oil.  For the suspensions, a linear fit yields a slope of $0.02ms^{-1}$ (blue) for times $t > 2ms$.}
	\label{self similar}
\end{figure}

Our capillary break-up experiments are performed in a home-made capillary extensional break-up rheometer (CaBER) device. A well defined quantity of the sample liquid is placed with a pipette between two steel discs with a diameter of $2mm$. The lower disc is fixed and the upper disc can be drawn apart very slowly with a linear motor (P01-23x80, Linmot, Spreitenbach, Switzerland). To measure the minimal neck diameter $h(t)$ as a function of time, a LED lamp is used to produce a shadowgraph image of the capillary bridge. Light that passes through the center of the filament is not diffracted and produces a bright line in the middle of the filament, while the light that interacts with the boundaries of the filament gets diffracted so they appear dark. In principle, the intensity profile of the diffracted light can be evaluated to achieve spatial resolutions below $100nm$ \cite{Super}. The filament is filmed by a $10 bit$ high speed camera (X-Stream XS-5, IDT, Tallahassee, USA) at up to $8.2 kHz$ frame rate depending on the spatial resolution. The camera has $1280\times1024$ pixels with a size of $12\times12\mu m$. The camera is equipped with a tube lens and microscope objectives (Nikon) with magnifications ranging from $2$ to $10$ fold.

Our samples have been described before \cite{vanDeen13}. In brief, spherical polystyrene beads Dynoseeds from Microbeads with grain diameters $d = 20, 40$ and $80 \mum$ at volume fractions $\phi  = V_g/V_0= 1 $ to $  3 \%$, with $V_g$  the volume of grains and $V_0$ the total volume, are dispersed in a silicon oil (Shin Etsu SE KF-6011) closely matching their density. The viscosity of the pure oil is $\eta_0 = 0.18$ Pa.s and its surface tension $\sigma = 21\pm1$ mN/m at $T = 21^\circ$ C. The volume fractions were chosen such that in most experimental runs only single beads would be present in the capillary bridge.

At these low volume fractions, the viscosity of the suspensions is well described by Einstein´s formula $\eta=\eta_0(1+2.5 \phi)$. The maximum difference between the viscosity of the suspension and the interstitial fluid is between  $2.5\%$ or $7\%$ and this small change in viscosity has negligible effects on the thinning dynamics of the viscous thread \cite{vanDeen13}.

A typical series of images close to pinch-off for a sample with $80 \mum$ beads at $\phi = 3 \%$ is shown in Fig. \ref{frames}(top). One observes that single beads remain in the thinning thread, leading to perturbations of the free surface and a larger curvature. The distribution of the beads becomes inhomogeneous as the thread thins further and at some moment the thinning dynamics become localized between two beads and finally the capillary bridge breaks between the two beads that are furthest apart. In the following we will show that the exact thinning dynamics of this regime depend on the particle size and the individual distribution of beads in the thread.

\begin{figure}[h]
	\centering
		\includegraphics[width=1.0\linewidth]{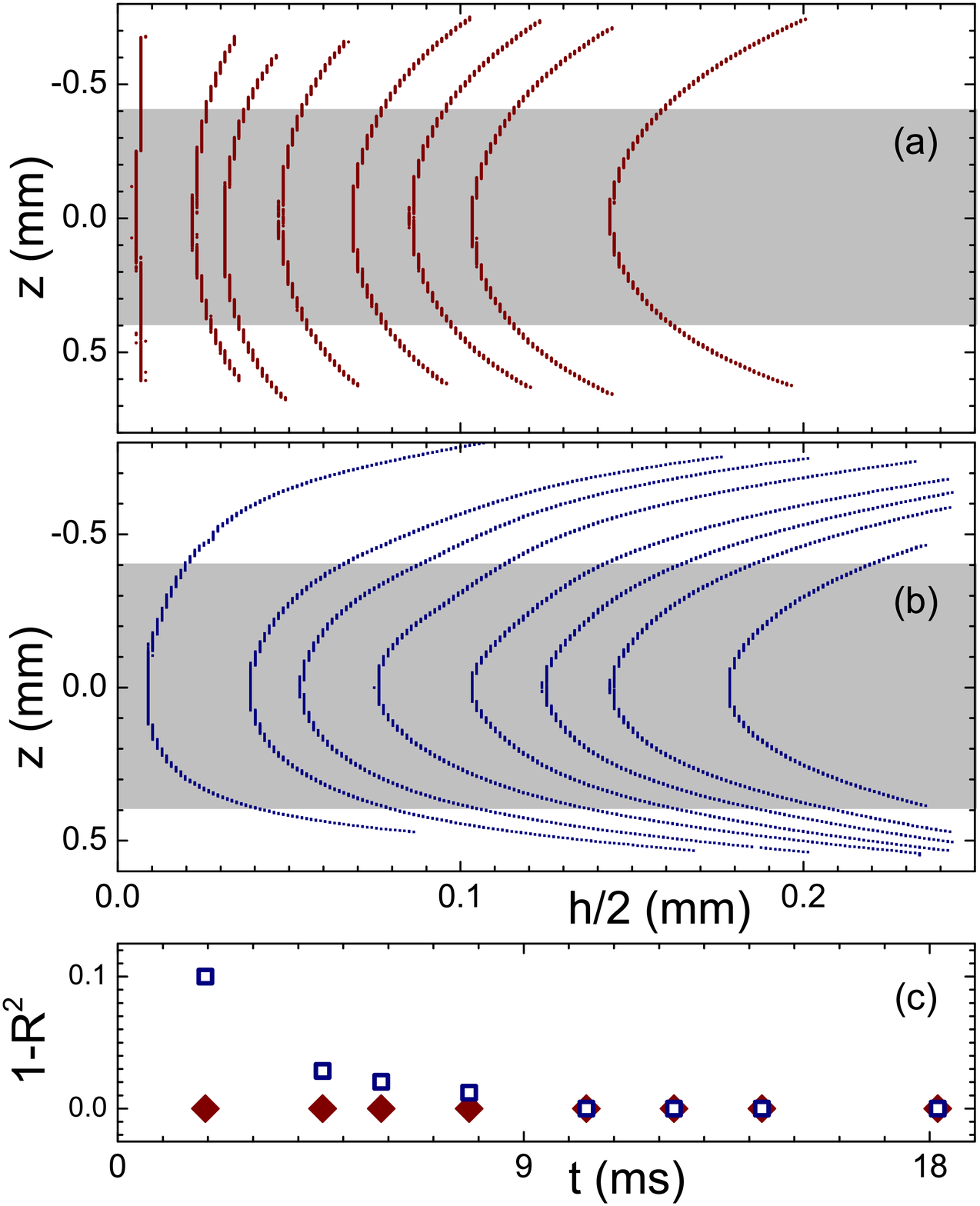}
	\caption{(Color online) Shapes of the thinning capillary filaments. The $z$ axis represents the vertical direction with gravity being oriented from left to right.  (a) For the oil (b) for the suspension with $80 \mum$ beads $(\phi=3\%)$ (b). The shapes corrospondend to times $t$ = $18.18$, $14.28$, $12.33$, $10.39$, $7.79$, $5.84$, $4.54$, and $1.94$ ms.(c) One minus the correlation coefficient (the residual sum of squares) of a parabolic fit of the shapes, indicating the symmetric shape for the pure oil (rhombi) and asymmetric for the suspension (squares).}
	\label{profiles}
\end{figure}

If we define the time $t=\tau_p-\tau$ as the difference of the actual time $\tau$ from the moment of pinch-off $\tau_p$  and plot the minimum diameter of the capillary bridge $h(t)$ as a function of time for the pure silicon oil (Fig. \ref{case studies}, black circles) we recognize the well known different regimes of capillary thinning. First, for times $ t >> 10$~ms before pinch-off, the dynamics are still governed by the exponential growth of the primary Rayleigh Plateau instability (data not fully shown). Approaching pinch-off, the system will then first follow the viscosity dominated Stokes (S) self similar law \cite{Eggers1993}

\begin{equation}
h(t) =0.0709 \frac{\sigma}{\eta} \times t= 0.015 \frac{m}{s} \times t.
\end {equation}

 With higher temporal and spatial resolution  (black circles in Fig. \ref{self similar}), a further transition to the Navier-Stokes (NS) self similar regime with
\begin{equation}
h(t) =0.0304 \frac{\sigma}{\eta} \times t= 0.007 \frac{m}{s} \times t
\end {equation}

 is observed at $t_{NS} \approx 2ms$ corresponding to a minimal diameter of $h_{NS} \approx 14\mum$.


For the suspensions (Fig. \ref{case studies}, black triangles) the slopes of the Stokes regime are identical to the slopes for the pure oil, confirming our hypothesis that at the small volume fractions we work with, the suspension viscosity can be considered to be identical to the viscosity of the interstitial fluid. However, the capillary thread does not evolve directly from the Stokes to the Navier-Stokes regime regime for the suspensions, but we observe an intermediate accelerated regime with a higher slope compared to the Stokes regime (Fig. \ref{case studies}). The average slope of this accelerated regime for all our experimental runs is $0.022 \pm 0.002$ m/s. No significant dependence on the particle size is observed within our experimental resolution, but the slope of a given experiment depends on the distribution of particles within the filament as we will show below.

At even later stages of the detachment the Navier-Stokes regime, with a slope that is identical to the pure oil, is recovered again (Fig. \ref{self similar}).

The Stokes and the accelerated regime are not only distinguishable by their different scalings of the minimum neck diameters but also by their symmetries along the z-axis. The shape of the thread in the Stokes regime is z-symmetric around the minimum diameter \cite{Eggers1993}, as is shown for the pure oil in Fig. \ref{profiles}~a. In contrast, at the same minimum filament diameter, the filament shapes of the suspension have not only a much larger curvature but they evolve also from a symmetric shape in the Stokes regime to a highly asymmetric shape in the accelerated regime (Fig. \ref{profiles}~b). This is quantitatively shown by representing the correlation coefficient (the residual sum of squares) of a quadratic fit to the filament shape (Fig. \ref{profiles}~c). The asymmetric profile of the thread also proves that the accelerated regime is not a transient inertial regime that has recently been observed in Newtonian fluids \cite{Barsan2015} and where the profile is symmetric. However, it is the specific shape of the filament of the suspension with its large curvature that leads to the accelerated regime and we will describe in the following how the size and distribution of the particles affect this regime.

\begin{figure}[h]
	\centering
		\includegraphics[width=1\linewidth]{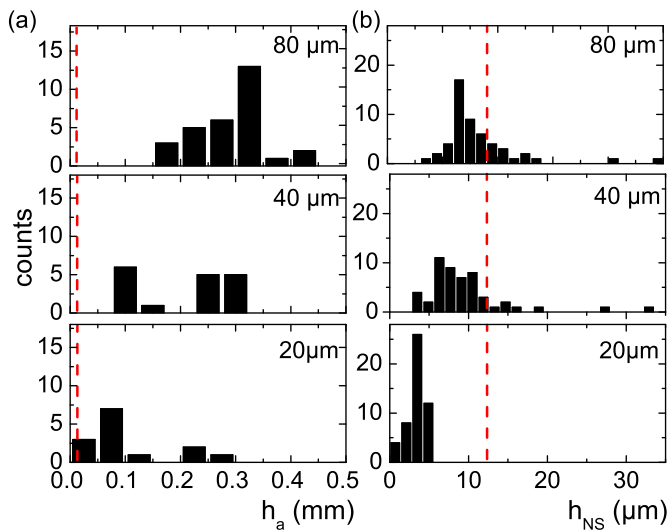}
	\caption{(Color online) a) Histogram of transition diameters $h_{a}$ for the suspensions from Stokes to the accelerated regime. The dashed line indicates the transition diameters $h_{NS}$ at the transition from the Stokes to the Navier-Stokes regime for the pure oil. b) Histogram of the distribution of transition diameters from the accelerated regime to the Navier Stokes regime $h_{NS}$. The dashed line indicates the transition diameters $h_{NS}$ at the transition from the Stokes to the Navier-Stokes regime for the pure oil.}
	\label{histograms}
\end{figure}

 \begin{figure}[htbp]
	\centering
		\includegraphics[width=0.9\linewidth]{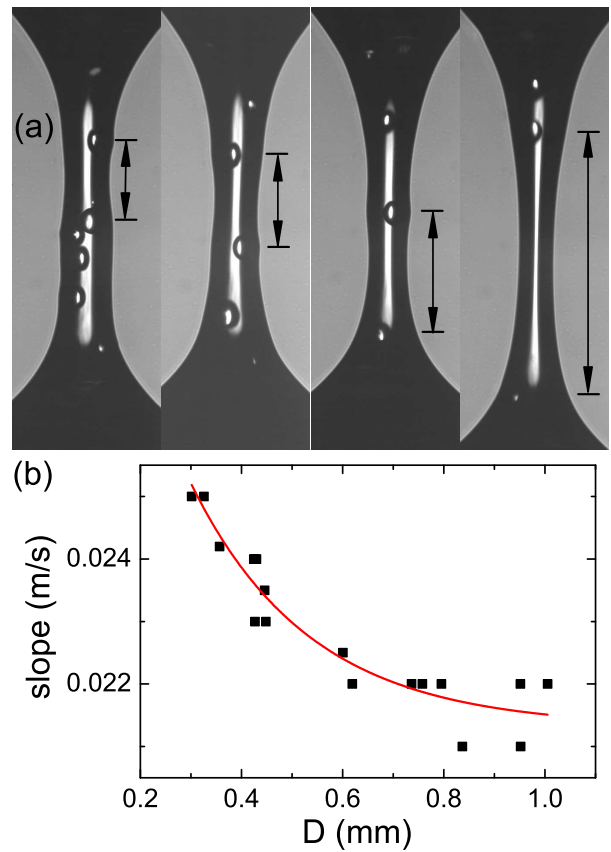}
	\caption{(Color online) a) Highspeed images for different experimental runs at $\approx7$ ms before pinch-off (accelarated regime). The arrows indicate the distance $D$ between the two beads where final break-up will occur. The slope of the minimum diameter vs. time curve $h(t)$ are from left to right: $0.025$, $0.024$, $0.023$ and $0.022$ mm/ms. b) The slope of the accelerated regime for the $80 \mum (\phi=3\%)$ suspension as a function of the distance between the two beads that are farthest apart in the filament at $t =7 ms$ before pinch-off (cp. Fig. 1). The line is a guide for the eye only.}
	\label{slope}
\end{figure}

The minimal neck diameters at which the crossover to the accelerated regime and subsequently to the Navier-Stokes regime occur are function of the particle size, as can already be seen from Fig. \ref{case studies}. This indicates that the transition is caused by the perturbation of the thread due to the presence of beads (see Fig. \ref{frames}). To quantify this further, we have measured the minimal neck diameters at these transition points for a total of $236$ experiments for the three different particle diameters. Histograms of the minimal diameters at the transition to the accelerated regime are shown in figure \ref{histograms}~a. It is clearly observed that for smaller bead sizes the transition takes place at smaller neck diameters. For the smallest bead size the transition to the accelerated regime sometimes overlaps with the transition to the Navier-Stokes regime for the pure oil. The increase of the transition diameter with particle size is in agreement with previous observations by Bonnoit \al \cite{Bonnoit12} who obtained average transition diameters comparable to those reported here. They also showed that the dependence of the average transition diameter on the particle size does not follow a simple scaling law and seems to be slightly less than linear. Here we show that there is a large variation in transition diameter for a given bead size, as can be seen from the histograms of Fig. \ref{histograms}. We will show below that this large variation can be explained by the varying distribution of individual particles in the viscous thread close to pinch-off.  Fig. \ref{histograms}~b shows the minimal neck diameters at the crossover from the accelerated to the Navier-Stokes regime.  Once more, the transition diameters $h_{NS}$ become smaller with decreasing particle diameter and they are always smaller than the transition diameter $h_{NS}$ between the Stokes and the Navier-Stokes regime for the pure oil (Fig. \ref{histograms}~b). As before, the transition diameters $h_{NS}$ show variation for a given particle size due to the variation in particle distributions in the thread.

 We will now quantify the acceleration caused by the particles. To do so we have measured the slope in the accelerated regime and have linked it to the largest distance between two particles in the viscous thread (Fig. \ref{slope}). In the accelerated regime the filament thinning is localized between these two particles and the viscous thread will break there. We chose to measure this distance at 7~ms before the pinch-off  where the thinning dynamics are well into the accelerated regime. We have performed this analysis for the largest particle size leading to the highest precision when determining particle positions in the thread. If we look at the snapshots at $t = 7$~ms we find that the filament will always break between the two particles that are most apart and that the slope of the $h(t)$ curve is a function of the distance $D$ of these two particles (Fig. \ref{slope}). The smaller this distance is, the larger is the curvature and the dynamics will be fastest. Interestingly, we find that the accelerated regime is observed even if the beads are only at the ends of the capillary thread, in agreement with observations by van Deen \al \cite{vanDeen13}. The thinning dynamics in the accelerated regime depend thus strongly on the microscopic distribution of the particles. The varying particle distribution is most likely also at the origin of the large variations in the values of the minimum neck diameter at the crossover to and from the accelerated regime. Note that a theoretical approach by Hameed and Morris \cite{Hameed2009} predicts a slowing down of the thinning of a viscous thread in the presence of particles, but in their analysis the particles were held fixed contrary to the experimental situation, probably at the origin of the observed differences.

In conclusion we have shown that the final stages of pinch-off in a very dilute suspension are accelerated compared to the pure suspending fluid, despite the fact that the bulk viscosity of the suspensions is nearly unchanged by the presence of particles at these small concentrations.  We show that the filament thinning dynamics for the suspensions evolve through a self-similar regime identical to the one of the intersticial fluid before entering an accelerated regime. At even later stages of the detachment process the suspensions recover the self-similar solution of the interstitial fluid and the thinning dynamics become identical to the interstitial fluid again.  The acceleration is caused by the perturbaion of the thread by the particles in the viscous thread. They locally introduce a curvature that is too large to accommodate the self similar solution of the Stokes regime. This leads to the strong acceleration of the thinning dynamics until the self similar solution is recovered and the thinning dynamics become identical to the pure fluid again.

The local distribution of isolated particles captured in the thread determines the filament diameter at which the crossover to the accelerated regime takes place. As the exact distribution varies from one experimental run to another a large variation is observed for the crossover diameters. The average value is function of the particle diameter, but no simple scaling between the crossover filament diameter and the particle size has been found. Similar observations are made for the filament diameter at which the accelerated regime crosses over to the self-similar solution again. The slope of the accelerated regime is also a direct consequence of the distribution of particles in the thread and we have shown that it is function of the distance between the last two remaining particles in the filament.

We have thus shown that the presence of isolated particles in the viscous thread strongly affect the thinning dynamics close to final pinch-off. The exact evolution of the shape of the viscous thread in time is function of the particle size and distribution in the thread.

\acknowledgments
 JF thanks the Alexander von Humboldt foundation and Global Site S.L.

\end{document}